\documentclass[preprint2]{emulateapj}
\usepackage{epstopdf}

\slugcomment{Accepted to ApJ}

\shorttitle{Nonmetastable Ammonia Masers in NGC~7538}
\shortauthors{Hoffman}

\begin{document}

\title{New Maser Emission from Nonmetastable Ammonia in NGC~7538. III. Detection of the (10,6) Transition and a Velocity Gradient}

\author{Ian M.\ Hoffman}
\affil{Wittenberg University, Springfield, OH 45501}
%\affil{Department of Physics, Wittenberg University, P.O.\ Box 720, Springfield, OH 45501}
\email{ihoffman@wittenberg.edu}

\begin{abstract}
We present the first astronomical detection of the $^{14}$NH$_3$ $(J,K) = (10,6)$ line: nonthermal emission at several velocities in the Galactic star-forming region NGC~7538.
Using the VLA we have imaged the (10,6) and (9,6) ammonia masers at several positions within NGC~7538 IRS~1.
The individual sources have angular sizes $\lesssim 0.1$~arcseconds corresponding to brightness temperatures $T_B \gtrsim 10^6$~K.
We apply the pumping model of Brown \& Cragg, confirming the conjecture that multiple ortho-ammonia masers can occur with the same value of $K$.
The positions and velocities of the (10,6) and (9,6) masers are modeled as motion in a possible disk or torus and are discussed in the context of recent models of the region.
\end{abstract}


\section{Introduction}

In an ongoing series of papers (Hoffman \& Kim 2011a, hereafter Paper~I; Hoffman \& Kim 2011b, hereafter Paper~II) we have examined the ammonia masers in NGC~7538 with the hope of developing new constraints on the poorly understood phenomenon of nonthermal emission from nonmetastable ($J>K$) states.
Although thermal emission from metastable ($J=K$) states is well developed into a method of thermometry of the emitting volumes ({\it e.g.}, Maret et al.\ 2009), nonmetastable transitions have received comparatively little attention.
Indeed, nonmetastable ammonia masers were discovered serendipitously in 1984 by Madden et al.\ (1986) after which only one emitting region (W51) was studied in depth ({\it e.g.}, Wilson, Johnston, \& Henkel 1990; Pratap et al.\ 1991) culminating in a pump model by Brown and Cragg (1991, hereafter BC91) that has never been tested.
With the exception of the discovery of a new maser source by Walsh et al.\ (2007), there has been no study of nonmetastable ammonia masers for 20 years.

Nonmetastable ammonia masers are found in Galactic high-mass star-forming regions.
Only five masering regions are known: NGC~7538, W51, DR21(OH), W49, and NGC~6334I.
Although a large-scale, dedicated survey for nonmetastable ammonia masers has never been conducted, the HOPS survey by Walsh et al.\ (2011) included the frequencies of the $(J,K)=(11,9)$ and (8,6) $^{14}$NH$_3$ transitions.
With a detection threshold of $\approx 2$~Jy (suitable to detect only the strongest known ammonia masers), the HOPS survey resulted in the detection of only one new source of possible ammonia-maser emission despite surveying more than 500 Galactic sources of water-maser emission.
In contrast, the sensitive ($\approx 100$~mJy) survey of 17 star-forming regions conducted by Madden et al.\ (1986) resulted in four detections.
Thus, the prevalence and diagnostic utility of ammonia masers in the study of star formation remain open questions.

\begin{figure*}[t]
\centering
%\epsscale{.80}
\includegraphics[angle=270,scale=0.60]{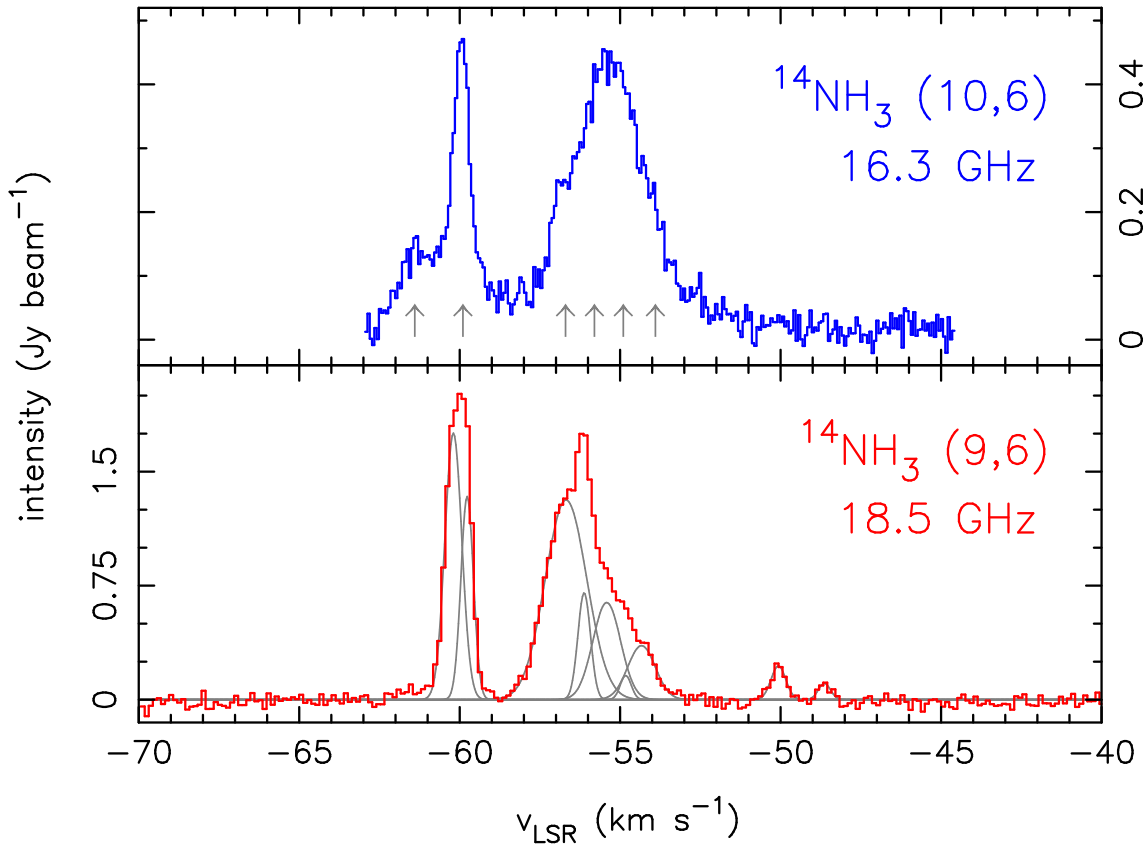}
%\plotone{f1.eps}
\caption{
VLA spectra for {\it (top)} $^{14}$NH$_3$ (10,6) and {\it (bottom)} (9,6) emission from NGC~7538 IRS~1.
Both data sets have had the contribution from continuum emission subtracted in the $u,v$ plane.
For display in this figure, each of these data sets has been convolved with the beam from Paper~I: $4\farcs6 \times 1\farcs9$ at a position angle of $34\degr$.
The gray vertical arrows in the (10,6) spectrum denote the velocity features whose sky positions are plotted in Figure~\ref{106}.
The gray curves in the (9,6) spectrum represent the fitted Gaussian functions that are summarized in Table~\ref{t1} and whose sky positions are plotted in Figure~\ref{96}.
\label{spectra}}
\end{figure*}

Brown and Cragg (BC91) developed a pumping scheme for ortho-ammonia ($K=3n; n=0,1,2,\ldots$) in which there is a relationship among the states having the same azimuthal angular momentum $K$ but different total angular momentum $J$.
These ``rungs'' of value $J$ within a ``ladder'' of value $K$ are suggested to be excited simultaneously, resulting in emission from multiple rungs for certain physical conditions.
Astronomically, only masering in isolated rungs -- such as (6,3) without accompanying emission in (5,3), (7,3), or elsewhere in the $K=3$ ladder -- has ever been observed from sources.
One exception may be NGC~6334I in which Walsh et al.\ (2007) report possible maser emission in (10,9) and (9,9) accompanying the known maser in (11,9).
In the laboratory, Willey et al.\ (1995) have pumped an isolated (4,3) maser.
The suggested interrelationship of the ladder rungs is important for two reasons:
(1) many of the ammonia masers in existence may yet be undetected since the frequencies of most rungs' transitions lie outside the range of traditional radio astronomy receivers, and
(2) the conditions for simultaneous excitation of multiple transitions are far tighter than for one transition and, therefore, more valuable as diagnostics of stellar environments.

With the study described in this paper, we have set out to improve both the frequency coverage and angular resolution of the study of nonmetastable ammonia masers.
With the advent of the upgraded Karl G.\ Jansky Very Large Array (VLA) of the NRAO\footnote{The National Radio Astronomy Observatory is a facility of the National Science Foundation operated under cooperative agreement by Associated Universities, Inc.}, receivers are now available at the frequencies of many transitions of interest.
Furthermore, improving upon the angular resolution of single dishes with interferometric imaging provides three significant benefits to understanding the excitation mechanism:
(1) the maser positions may be determined relative to the continuum emission in the region in order to constrain the pumping environs,
(2) the angular sizes of the masering regions may be constrained, providing for tighter limits on the brightness temperature of the emission and on the amplification gains, and
(3) sufficiently precise angular positions of different transitions and velocity components will aid in discriminating isolated from coincident pumping.
In addition, the general benefit of any maser study is that the kinematics of the gas in the region can be determined at the scale of $\sim 100$~AU.
With a deeper understanding of these basic parameters, nonmetastable ammonia masers can be built into a robust tool for understanding star formation.

\section{Observations and Results}

\subsection{Ku-band VLA `C' Configuration\label{106results}}

On 2012 January 31 we observed the $^{14}$NH$_3$ (10,6) ammonia transition at $\nu_{\rm rest} = 16.31932$~GHz, employing the 15 VLA antennas retrofitted with new $Ku$-band receivers at the time.
The relative separations between the usable antennas ranged from 0.1 to 2.6~km, resulting in a circular synthesized beam of FWHM diameter 1.8~arcseconds.
The flux, phase, and bandpass calibrators used were 3C48, 2322+509, and 3C147 with approximately 17 minutes of integration on the IRS~1 target.
We recorded a single circular polarization over a 1-MHz bandwidth centered in frequency on $v_{\rm LSR}=-53.8\,{\rm km}\,{\rm s}^{-1}$ divided into 256 spectral channels resulting in a velocity resolution of $0.1\,{\rm km}\,{\rm s}^{-1}$ and a total velocity coverage of approximately $18\,{\rm km}\,{\rm s}^{-1}$.
The data were reduced using AIPS.\footnote{The Astronomical Image Processing System is documented at \url{http://www.aips.nrao.edu/}.}
The {\it rms} background noise in a channel image is $17\,{\rm mJy}\,{\rm beam}^{-1}$, consistent with instrumental expectations.

\begin{figure}
\centering
%\epsscale{.80}
\includegraphics[angle=270,scale=0.40]{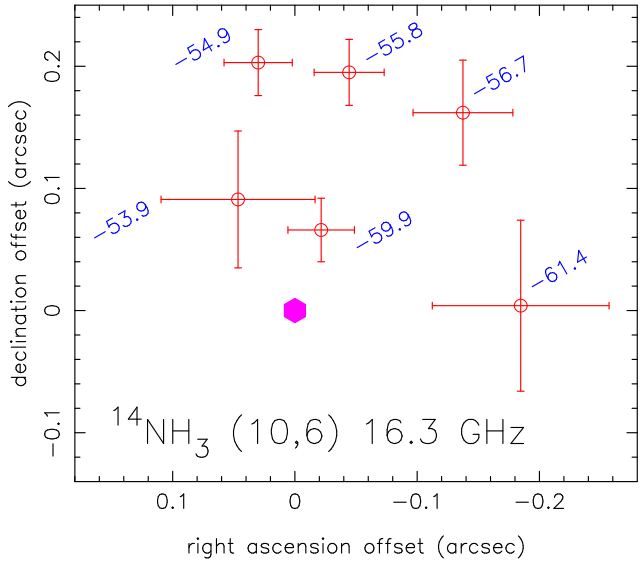}
%\plotone{spectrum.eps}
\caption{
Positions of the $^{14}$NH$_3$ $(10,6)$ masers in NGC~7538 IRS~1 from the VLA `C' configuration observations of 2012 January 31.
The synthesized beam is circular with a FWHM diameter of 1.8~arcseconds.
The diagonal labels are the velocities of the features denoted in Figure~\ref{spectra} and summarized in Table~\ref{t1}.
For the location of the maser in each imaged spectral channel, the error bars are the position uncertainty of a fitted two-dimensional Gaussian model.
The filled hexagon is the location of the peak of the 16.3-GHz continuum emission from which the position offsets in this data set are measured.
As discussed in Section~\ref{comp}, these error bars are appropriate for determining the relative separation of the features but do not represent the precision of the absolute position registration.
\label{106}}
\end{figure}

In the top panel of Figure~\ref{spectra} is displayed the ammonia spectrum.
Due to spatial and spectral blending, Gaussian components could not be fit reliably to the velocity profile.
Nevertheless, obvious velocity features (denoted by vertical arrows in Figure~\ref{spectra}) were chosen subjectively in order to select the channel images from which the positions in Figure~\ref{106} were determined.
None of the velocity components are coincident on the sky, indicating that the emission arises in several distinct kinematic volumes.

Most of the masers are not angularly resolved in the VLA observations and have an upper limit on deconvolved size of approximately 1 arcsecond.
These sizes correspond to a lower limit on brightness temperature of $T_B > 2000\,{\rm K}$.
The maser component at $v_{\rm LSR} = -59.9\,{\rm km}\,{\rm s}^{-1}$ has a resolved size of 40~milliarcseconds corresponding to a brightness temperature of $T_B = 1.1 \times 10^6$~K.
Assuming that this maser amplifies the 16.3-GHz background continuum emission from IRS~1 following $T_B = T_{bg}e^{-\tau}$, we find the gain to be $\tau = -4.3$.
The measured parameters for all of the (10,6) emission is summarized in Table~\ref{t1}.

\subsection{K-band VLA `B' Configuration\label{96results}}

On 2012 June 15 we observed the $^{14}$NH$_3$ (9,6) ammonia transition at $\nu_{\rm rest} = 18.49939$~GHz, employing all 27 antennas of the VLA.
The relative separations between the antennas ranged from 0.3 to 11.1~km, resulting in a synthesized beam of $0\farcs33 \times 0\farcs26$ at a position angle of $-15\degr$; the array was not sensitive to image features larger than approximately 10~arcseconds.
The flux, phase, and bandpass calibrators used were 3C48, 2322+509, and 3C147 with approximately 45 minutes of integration on the IRS~1 target.
We recorded a single circular polarization and a 2-MHz bandwidth centered in frequency on $v_{\rm LSR}=-53.8\,{\rm km}\,{\rm s}^{-1}$ divided into 256 spectral channels resulting in a velocity resolution of approximately $0.2\,{\rm km}\,{\rm s}^{-1}$ and a total velocity coverage of approximately $32\,{\rm km}\,{\rm s}^{-1}$.
The data were reduced using AIPS.
The {\it rms} background noise in a channel image is $14\,{\rm mJy}\,{\rm beam}^{-1}$, consistent with instrumental expectations.

\begin{figure}
\centering
%\epsscale{.80}
\includegraphics[angle=270,scale=0.50]{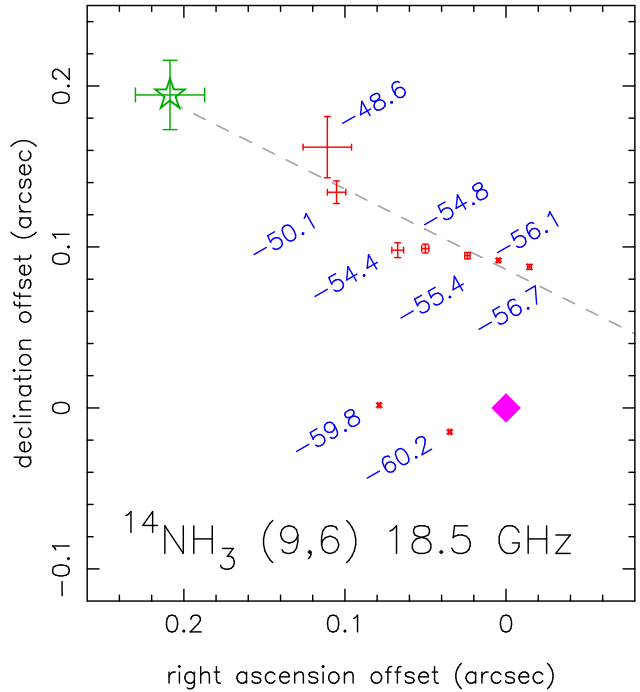}
%\plotone{spectrum.eps}
\caption{
Positions of the $^{14}$NH$_3$ $(9,6)$ masers in NGC~7538 IRS~1 from the VLA `B' configuration observations of 2012 June 15.
The images have a synthesized beam of $0\farcs33 \times 0\farcs26$ at a position angle of $-15\degr$.
The diagonal labels are the velocities of the features summarized in Table~\ref{t1} and plotted in Figure~\ref{spectra}.
For the location of the maser in each imaged spectral channel, the error bars are the position uncertainty of a fitted two-dimensional Gaussian model.
The filled diamond is the location of the peak of the 18.5-GHz continuum emission from which the position offsets in this data set are measured.
As discussed in Section~\ref{comp}, these error bars are appropriate for determining the relative separation of the features but do not represent the precision of the absolute position registration.
The gray dashed line denotes the direction of the Keplerian gradient discussed in Section~\ref{gradient} and displayed in Figure~\ref{kepler}.
The star symbol is the fitted location of the central object of the Keplerian model, with error bars denoting formal fitting errors.
\label{96}}
\end{figure}

Using the VLA, we detect the same velocity components as described in Paper~II for the observations using the Robert C.\ Byrd Green Bank Telescope (GBT), with the exception of $v_{\rm LSR} = -46.7\,{\rm km}\,{\rm s}^{-1}$ which is below the detection threshold of the current observations.
In the bottom panel of Figure~\ref{spectra} is displayed the VLA ammonia spectrum and fitted velocity components, which are summarized in Table~\ref{t1}.
Due to spatial and spectral blending, the fits are not reliable to the widths and amplitudes of the components having $-57\,{\rm km}\,{\rm s}^{-1} > v_{\rm LSR} > -53\,{\rm km}\,{\rm s}^{-1}$.
Nevertheless, the center velocities of the features are apparent and were used to select the channel images from which the positions in Figure~\ref{96} were determined.

The current results are also consistent with the VLA observations presented in Paper~I, with the maser emission near $v_{\rm LSR} = -60.2\,{\rm km}\,{\rm s}^{-1}$ lying southeast of the peak of continuum emission and of the masers near $v_{\rm LSR} = -57\,{\rm km}\,{\rm s}^{-1}$.
Furthermore, all of the velocity components have positions at the location of IRS~1 in NGC~7538.
These results are consistent with the GBT observations presented in Paper~II and allow us to rule out the association of these ammonia masers with any other sources in the region (such as IRS~9 or MM2).

The velocity feature near $v_{\rm LSR} = -57\,{\rm km}\,{\rm s}^{-1}$ is far less pronounced in the current spectrum than in previous observations, indicating some variability among the several blended components between $-57\,{\rm km}\,{\rm s}^{-1} > v_{\rm LSR} > -53\,{\rm km}\,{\rm s}^{-1}$.
In addition, the entire amplitude scale is $\approx 2$ times larger in the current paper than in Papers I and II.
The flux scale of the current observations was confirmed independently against both 3C48 and 3C147 with consistent results, leaving the discrepancy an open question; we hesitate to conclude that all of the maser components varied in unison by a factor of two since multiplicative scaling is a common instrumental uncertainty.

As summarized in Table~\ref{t1}, many of the masers are angularly resolved, having sizes on the order of 100~milliarcseconds ($\approx 300$~AU at a distance of 2.65~kpc, Moscadelli et al.\ 2009).
These brightness temperatures of $T_B \sim 10^6$~K correspond to unsaturated amplification gains of $\tau \approx -5$ of the background continuum emission.

As reported in Papers I and II, the velocity components near $v_{\rm LSR} = -60\,{\rm km}\,{\rm s}^{-1}$ have been relatively unchanged since their discovery in 1984 (there are unpublished reports of variability by a factor of 2.5 in the year following discovery, see Schilke et al.\ 1991).
In contrast to these older masers, all of the redward emission that has arisen subsequently is found to lie along the same northeast-southwest line (see Fig.~\ref{96}).
This velocity distribution is discussed in Section~\ref{gradient}.

\begin{deluxetable}{l r@{.}l r@{.}l r r@{.}l r@{.}l}[b]
\tablecolumns{10}
\tablewidth{0pt}
\tablecaption{Fitted Properties of Ammonia Emission\label{t1}}
\tablehead{
\multicolumn{1}{c}{$v_{\rm LSR}$} & \multicolumn{2}{c}{$I$} & \multicolumn{2}{c}{$\Delta{v}_{\rm FWHM}$} & \multicolumn{1}{c}{$\theta_{\rm maser}$} & \multicolumn{2}{c}{$T_B$} & \multicolumn{2}{c}{$\tau$} \\
\multicolumn{1}{c}{(${\rm km}\,{\rm s}^{-1}$)} & \multicolumn{2}{c}{(${\rm Jy}\,{\rm bm}^{-1}$)} & \multicolumn{2}{c}{(${\rm km}\,{\rm s}^{-1}$)} & \multicolumn{1}{c}{(mas)} & \multicolumn{2}{c}{($10^6$~K)} & \multicolumn{2}{c}{}
}
\startdata
\multicolumn{10}{c}{$^{14}$NH$_3$ (10,6)} \\ \tableline
$-$61.4\tablenotemark{a} & \multicolumn{2}{c}{\tablenotemark{b}} & \multicolumn{2}{c}{\tablenotemark{b}} & $<$1000 & $>$0&001 & $<-$1&1 \\
$-$59.95(1) & 0&37(1) & 0&55(3) &     40 &   1&1 &   $-$4&3 \\
$-$56.7\tablenotemark{a} & \multicolumn{2}{c}{\tablenotemark{b}} & \multicolumn{2}{c}{\tablenotemark{b}} & $<$800  & $>$0&002 & $<-$1&7 \\
$-$55.8\tablenotemark{a} & \multicolumn{2}{c}{\tablenotemark{b}} & \multicolumn{2}{c}{\tablenotemark{b}} & $<$900  & $>$0&003 & $<-$2&0 \\
$-$54.9\tablenotemark{a} & \multicolumn{2}{c}{\tablenotemark{b}} & \multicolumn{2}{c}{\tablenotemark{b}} & $<$600  & $>$0&005 & $<-$2&8 \\
$-$53.9\tablenotemark{a} & \multicolumn{2}{c}{\tablenotemark{b}} & \multicolumn{2}{c}{\tablenotemark{b}} & $<$700  & $>$0&002 & $<-$1&9 \\ \tableline
\multicolumn{10}{c}{$^{14}$NH$_3$ (9,6)} \\ \tableline
$-$60.20(4) & 1&78(8) & 0&63(6) &     70 &    1&3 &  $-$5&0 \\
$-$59.77(3) & 1&3(3)  & 0&45(3) &  $<$90 & $>$0&6 & $<-$4&3 \\
$-$56.7(5)  & \multicolumn{2}{c}{\tablenotemark{b}} & \multicolumn{2}{c}{\tablenotemark{b}} & 50 & 2&0\tablenotemark{b} & $-$5&5\tablenotemark{b} \\
$-$56.1(2)  & \multicolumn{2}{c}{\tablenotemark{b}} & \multicolumn{2}{c}{\tablenotemark{b}} & 60 & 0&7\tablenotemark{b} & $-$4&4\tablenotemark{b} \\
$-$55.4(8)  & \multicolumn{2}{c}{\tablenotemark{b}} & \multicolumn{2}{c}{\tablenotemark{b}} & 60 & 0&6\tablenotemark{b} & $-$4&3\tablenotemark{b}  \\
$-$54.8(7)  & \multicolumn{2}{c}{\tablenotemark{b}} & \multicolumn{2}{c}{\tablenotemark{b}} & $<$90 & $>$0&07\tablenotemark{b} & $<-$2&2\tablenotemark{b} \\
$-$54.3(9)  & \multicolumn{2}{c}{\tablenotemark{b}} & \multicolumn{2}{c}{\tablenotemark{b}} & 100 & 0&13\tablenotemark{b} & $-$2&9\tablenotemark{b}  \\
$-$50.07(3) & 0&22(2) & 0&54(6) & $<$150 & $>$0&04 & $<-$1&9 \\
$-$48.64(5) & 0&11(2) & 0&5(1)  & $<$200 & $>$0&01 & $<-$0&7 \\ \tableline
\enddata
\tablecomments{The number in parentheses is the uncertainty in the final digit. The fitted angular sizes of the masers have a precision of one significant digit, from which the brightness temperatures and gains are calculated.}
\tablenotetext{a}{Approximate velocity of feature.}
\tablenotetext{b}{Fit uncertain due to blending of features.}
\end{deluxetable}

\subsection{Comparison Between Observations\label{comp}}

The absolute position registration between the 16.3-GHz and 18.5-GHz observations is limited by interferometer phase calibrations to a precision of approximately 200 milliarcseconds.
Both observations contain an image of the continuum radiation from IRS~1: the fitted peak of the image of the 16.3-GHz continuum emission lies northwest (position angle $-61\degr$) of the fitted peak of the image of the 18.5-GHz continuum emission by 50 milliarcseconds.
We assume these peaks to be coincident in position and use them to register the two observations to each other: the maser positions plotted in Figures \ref{106} and \ref{96} are relative to the peak of the continuum emission at that frequency.
The relative location of the masers to each other in a single observation is far more precise ($\approx 5$~mas) than an unregistered comparison between observations.
The location of the peak of the continuum emission in the 18.5-GHz observations is $\alpha_{\rm J2000} = 23\,{\rm h}\ 13\,{\rm m}\ 45\fs364$, $\delta_{\rm J2000} = 61\degr\ 28\arcmin\ 10\farcs34$.

Sandell et al.\ (2009) discuss extensively the morphology of the continuum emission from IRS~1.
Based on observations with very high angular resolution, they describe differences in morphology at the 150-mas scale between 14.9 and 22.4~GHz.
Given the poorer angular resolution and smaller difference in frequencies of our current observations, we consider our assumed 50-mas registration to be sufficiently precise for our discussion.

Alternatively, we may assume coincident positions for (10,6) and (9,6) ammonia masers with comparable velocities (similar to the registration method used by Surcis et al.\ [2011] for water and methanol masers).
Most notably, the brightest spectral features at $v_{\rm LSR} \approx -59.9\,{\rm km}\,{\rm s}^{-1}$ and $v_{\rm LSR} \approx -56.7\,{\rm km}\,{\rm s}^{-1}$ have comparable relative separations at each frequency.
The possibility of spatially coincident maser emission from both (10,6) and (9,6) is addressed in the next section.

\section{Discussion}

\subsection{Pump Model}

Brown and Cragg (BC91) suggest a pumping scheme for the (6,3) maser in W51 that accounts for the lack of maser emission in the (5,3) and (4,3) rungs of the $K=3$ rotational ladder.
The model requires radiative transitions from the vibrational ground state that are driven by an adjacent infrared source; vibrationally excited ammonia has been detected in NGC~7538 (Schilke et al.\ 1990).
The subsequent rapid radiative decays back into the vibrational ground state would invert all rungs of the ladder were the populations not mitigated by competing collisions which thermalize some of the transitions.
For certain ranges of infrared excitation, ammonia column, collision frequency, and collision selectivity, an ortho-ammonia maser can be produced in a specific rung.
Although the pumping scheme is heavily dependent on the cross-sections for selective collisions between ammonia and molecular hydrogen, the BC91 model is found to be insensitive to the input of either the rates extrapolated from helium calculations or the rates calculated specifically for molecular hydrogen.
Furthermore, BC91 note that the conditions for masers with $K=6$ are much different than for $K=3$, and so these kinematic volumes in IRS~1 should be devoid of $K=3$ emission.
Indeed, the (6,3) and (4,3) transitions have been studied in NGC~7538 and neither show maser emission (Madden et al.\ 1986; Schilke et al.\ 1991).

In the laboratory, Willey et al.\ (1995) successfully generated a nonmetastable ortho-ammonia maser in the (4,3) transition (in isolation from [5,3] or [3,3] emission) using exclusively collisional pumping (see also, Schilke et al.\ 1991).
In contrast to the analysis by Brown and Cragg, the laboratory maser was generated only in collisions with helium and could not be produced under any conditions with molecular hydrogen.
Flower, Offer, and Schilke (1990) note that population inversion is expected to depend on the ratio of ortho-H$_2$ to para-H$_2$ participating in the collisions.
Willey et al.\ (2000) went on to quantify significant differences between NH$_3$-He collisions and NH$_3$-H$_2$ collisions.
Furthermore, the laboratory maser is explained using transitions from the $K=0$ ladder even though that process is ruled out of the BC91 analysis.
Nevertheless, all groups of authors recognize that the role of collisions in pumping is predicated upon having level- and parity-dependent selectivity.

As noted by BC91, all theoretical work on ammonia masers has been limited by the lack of astronomical studies of the rungs higher in energy than the masering rung.
In the current paper, we present the first detection of masers in adjacent rungs of an ortho-ammonia ladder and discuss here briefly the implications of simultaneous maser emission from (10,6) and (9,6).
We are currently undertaking observations toward a complete survey of the ortho-ammonia ladders in NGC~7538 and withhold a larger discussion of level populations for a subsequent paper.

Regardless of how the 16.3- and 18.5-GHz observations are registered (see \S\ref{comp}), there is a position coincidence of (10,6) and (9,6) maser emission at either $v_{\rm LSR} \approx -54\,{\rm km}\,{\rm s}^{-1}$ or at $-60\,{\rm km}\,{\rm s}^{-1}$ and $-57\,{\rm km}\,{\rm s}^{-1}$.
Brown and Cragg explicitly address the possibility of the same volume of gas supporting both population inversions simultaneously.
In the same way that they predict a (6,3) maser to be accompanied by (7,3) -- and possibly also higher values of $J$ -- we have found the (9,6) maser to be accompanied by (10,6).
Qualitatively, we confirm their conjecture that masers in a given ladder may not appear in isolation but rather in all rungs above a critical value of $J=9$.
Quantitatively, little can be concluded without the detailed cross-section information, especially the mitigation of the radiative decays that is necessary to produce these (10,6) and (9,6) results.
With future work, a sufficiently complete census of the rungs in the $K=6$ ladder will allow the most selective level- and parity-dependent NH$_3$-H$_2$ cross-sections to be deduced, even in the absence of laboratory results.

\subsection{Velocity Gradient\label{gradient}}

In terms of both ages and positions, the (9,6) masers with $v_{\rm LSR} \approx -60\,{\rm km}\,{\rm s}^{-1}$ are qualitatively different than the younger, redward masers having $v_{\rm LSR} > -58\,{\rm km}\,{\rm s}^{-1}$.
Indeed, the straight-line position arrangement of the redward masers is accompanied by a monotonic velocity gradient that is not shared by the older masers.
The northeast-southwest orientation of the gradient (see Fig.~\ref{96}) is consistent with the orientation of many other proposed disks in the region (see Beuther et al.\ [2012] for a recent review).

\begin{figure}
\centering
%\epsscale{.80}
\includegraphics[angle=270,scale=0.36]{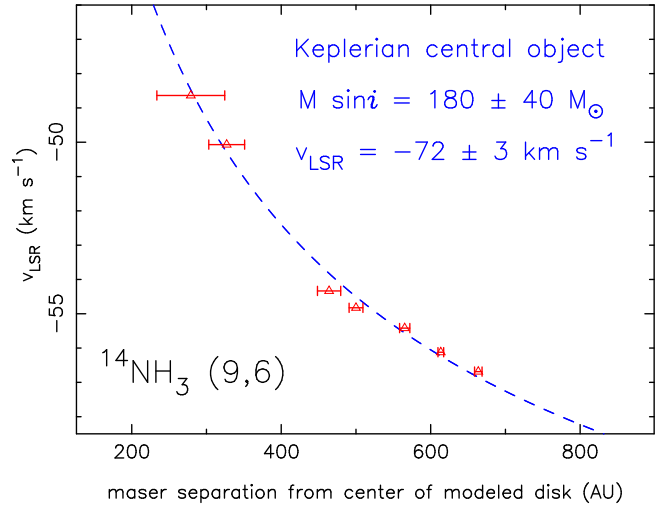}
%\plotone{spectrum.eps}
\caption{
One of the two interpretations of the velocity gradient: the Keplerian model discussed in Section~\ref{gradient}.
The data points are the masers whose positions lie along the gray line in Figure~\ref{96}, with error bars denoting uncertainty in position along that line.
The separations are measured from the fitted location of the central gravitating object, denoted with the star symbol in Figure~\ref{96}.
The conversion of angular position to the plotted linear distance assumes that the masers all lie with the central object in the plane of the sky at a distance of 2.65~kpc.
The dashed line is the fitted gradient model.
\label{kepler}}
\end{figure}

In our case, the gradient is well fit as Keplerian motion about an unseen source (the star symbol in Fig.~\ref{96}) that is significantly offset from the position of peak continuum emission.
Figure~\ref{kepler} shows the fit to the gradient that requires a massive central object ($180 \pm 40 M_\odot$) with a systemic velocity of $v_{\rm LSR} \approx -70\,{\rm km}\,{\rm s}^{-1}$.
The masers occur only in the southwest half of the disk where the rotating motion is away from the observer (counterclockwise as viewed from the northwest).
This sense of rotation is opposite the other proposed disks with a northeast-southwest plane.
The velocity of the fitted central object does not agree with the consensus velocity for IRS~1 ($v_{\rm LSR} \approx -57\,{\rm km}\,{\rm s}^{-1}$).
However, the fitted systemic velocity in our Keplerian model is comparable to that of the southern group of water masers described by Surcis et al.\ (2011) that lie approximately one arcsecond away from the fitted location shown in Figure~\ref{96}.

Although the fitted mass of the central object is unreasonably large in the Keplerian model, we hesitate to rule out this interpretation without understanding fully the contribution of the most redshifted masers to the model.
Indeed, the most redshifted maser described in Paper~II is not detected in the current observations.
Since the highest velocities have the most leverage on the fit, we cannot conclude that a Keplerian model is unphysical without improved positions for all of the masers having $v_{\rm LSR} > -51\,{\rm km}\,{\rm s}^{-1}$.
Nevertheless, this Keplerian model is only appropriate for the (9,6) masers, not the (10,6) masers.

\begin{figure}
\centering
%\epsscale{.80}
\includegraphics[angle=270,scale=0.35]{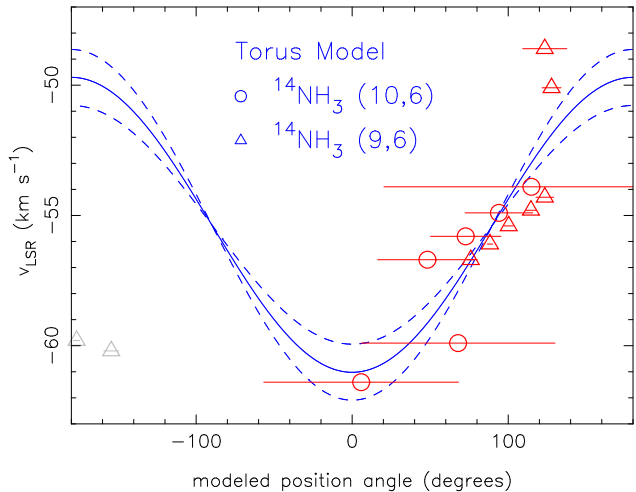}
%\plotone{spectrum.eps}
\caption{
One of the two interpretations of the velocity gradient: the torus model discussed in Section~\ref{gradient}.
The position angle is the azimuthal location of the masers in a suggested torus surrounding the central peak of continuum emission.
The circles represent (10,6) masers and the triangles represent (9,6) masers.
The two gray triangle markers with PA$\approx -160\degr$ are the ``older'' (9,6) masers that have been excluded from the model.
The horizontal error bars denote the uncertainty in the computed position angle due to the uncertainty in the sky positions of the masers.
The solid curve is the best-fit torus model, with a systemic velocity of $v_{\rm LSR} =-55.4\,{\rm km}\,{\rm s}^{-1}$ and a rotation speed of $5.7 \pm 1.1\,{\rm km}\,{\rm s}^{-1}$, with the dashed curves representing the upper and lower bounds of the fitted rotation speed.
As discussed in Section~\ref{gradient}, this modeling follows the analysis of Surcis et al.\ (2011) in the development of their Figure~9.
\label{torus}}
\end{figure}

As an alternative to a Keplerian interpretation, we consider that the (9,6) and (10,6) masers may mark the boundary of the torus/outflow region proposed by Surcis et al.\ (2011).
Those authors model the velocity gradients of methanol masers as arising in a torus with a flat rotation curve.
The positions and velocities of the redward (9,6) ammonia masers agree well with the methanol gradients (their Figures 4 and 9).
Furthermore, the distribution of (10,6) masers (Fig.~\ref{106}) can also be understood in this manner.
Using the parameters from Surcis et al.\ (see references therein) for the orientation of the torus (an outflow axis directed 40\degr\ west of north and an inclination of 32\degr), the azimuthal position angles in the torus can be computed for the ammonia masers.
We determine the position angles with the same sense and fiducial as presented in their Figure~9, and produce our analogous Figure~\ref{torus} for ammonia.
In fitting our data (the solid and dashed curves in Fig.~\ref{torus}), we find a systemic velocity of $v_{\rm LSR} =-55.4 \pm 0.6\,{\rm km}\,{\rm s}^{-1}$ and a rotation speed of $5.7 \pm 1.1\,{\rm km}\,{\rm s}^{-1}$, in good agreement with the quoted methanol parameters of $-57\,{\rm km}\,{\rm s}^{-1}$ and $7\,{\rm km}\,{\rm s}^{-1}$, respectively.
As discussed in Paper~II, ammonia exhibits a higher range of velocities in IRS~1 than any other maser species besides water, and the ammonia masers at the ends of the velocity range are critical to our fit to the torus model.
Although extraordinary for masers in the region, these ammonia velocities are nevertheless consistent with the range $-65\,{\rm km}\,{\rm s}^{-1} < v_{\rm LSR} < -45\,{\rm km}\,{\rm s}^{-1}$ quoted by Klaassen et al.\ (2009) for the bulk rotation of SO$_2$ gas in the torus. 

It is important to note that, in the torus model, all of the ammonia masers and some (group A) of the methanol masers lie in the part of the torus that is on the far side of the central object.
This scenario requires that the ionized outflow in the foreground at those locations be optically thin at frequencies as low as 6~GHz (which is not inconsistent with observations, {\it e.g.} Sandell et al.\ 2009).
Furthermore, the assumptions in Sections \ref{106results} and \ref{96results} that the masers amplify the free-free continuum emission from IRS~1 are not valid and the maser gains are much stronger (more negative) than quoted in Table~\ref{t1}.
Also, note that the (9,6) masers near $-60\,{\rm km}\,{\rm s}^{-1}$ have been excluded from the fit (the gray markers in Fig.~\ref{torus}).

One possible experiment to discriminate between a Keplerian interpretation and a torus interpretation of the ammonia gradient is to monitor the proper motion of the masers.
In the case of Keplerian motion, the innermost masers (the most redshifted) will complete an entire orbit in $\approx 50$~years, and therefore vary on few-year timescales by either moving across the sky or vanishing as the beaming and velocity coherence become unfavorable.
In the case of slow rotation with the torus, the masers would take approximately 200 years to move 100~milliarcseconds.

\section{Conclusion}

We present the first astronomical detection of the $^{14}$NH$_3$ $(J,K) = (10,6)$ line.
Using interferometric images of both the (10,6) and (9,6) masers in NGC~7538 IRS~1, we find maser amplification gains of $\tau \approx -5$ and brightness temperatures of $T_B \sim 10^6$~K.
Although many of the masers are isolated in both position and velocity, we note some possible coincidences between (10,6) and (9,6) emission which are discussed in the context of the pumping model of Brown and Cragg (1991).
All of the new (9,6) masers reported in Papers I and II are found to have a velocity gradient.
This gradient is consistent with both (1) Keplerian orbits about a very massive, unseen object and (2) rotation in a torus around the known peak of continuum emission.
The torus model can also be used to explain the positions of the (10,6) masers.
In either case, the ammonia masers are apparently valuable markers of the kinematics of the molecular environment.
Furthermore, the long-lived (9,6) masers originally discovered by Madden et al.\ (1986) do not fit well into either kinematic model and we leave open the possibility that they represent a different class than the other masers in this study.
In an ongoing effort to quantify the value of nonmetastable ammonia masers in the study of star formation, we are undertaking additional observations of other rungs in the rotational ladders with both single-dish and interferometric observatories.

\acknowledgments

This work is supported by the Lovejoy Science Fund of St.\ Paul's School and by the Weaver Fund of Wittenberg University.

{\it Facilities:} \facility{VLA ()}

\end{document}